\documentclass[universe,article,accept,pdftex,moreauthors]{Definitions/mdpi}

\firstpage{1}
\pubvolume{1}
\issuenum{1}
\articlenumber{0}
\pubyear{2023}
\copyrightyear{2023}
\datereceived{19 June 2023}
\daterevised{24 July 2023} 
\dateaccepted{28 July 2023}
\datepublished{ }
\hreflink{https://doi.org/} 

\Title{Constraining the Thickness of the Galactic Halo through Cosmic-Ray Anisotropy Using the Spatial-Dependent-Propagation Model}
\TitleCitation{Constraining the Thickness of the Galactic Halo through Cosmic-Ray Anisotropy Using the Spatial-Dependent-Propagation Model} 

\Author{Bing-Qiang 
Qiao $^{1}$, Yu-Hua Yao $^{1,2,}$*, Wei Liu $^{1}$, Qiang Yuan $^{3,4}$, Xiao-Jun Bi $^{1,5}$, Hong-Bo Hu $^{1,5}$ and Yi-Qing Guo $^{1,5,}$*}

\AuthorNames{Bing-Qiang Qiao, Yu-Hua Yao, Wei Liu, Qiang Yuan, Xiao-Jun Bi, Hong-Bo Hu and Yi-Qing Guo}

\AuthorCitation{Qiao, B.-Q.; Yao, Y.-H.; Liu,~W.; Yuan, Q.; Bi, X.-J.; Hu, H.-B.; Guo, Y.-Q.}

\address{%
$^{1}$ \quad Key Laboratory of Particle Astrophysics, Institute of High Energy Physics, Chinese Academy of Sciences, Beijing~100049,~China; qiaobq@ihep.ac.cn 
(B.-Q.Q.); liuwei@ihep.ac.cn~(W.L.); bixj@ihep.ac.cn~(X.-J.B.); huhb@ihep.ac.cn~(H.-B.H.)\\
$^{2}$ \quad College of Physics, Chongqing University, Chongqing~401331,~China\\
$^{3}$ \quad Key Laboratory of Dark Matter and Space Astronomy, Purple Mountain Observatory, Chinese Academy of~Sciences, Nanjing~210008,~China; yuanq@pmo.ac.cn\\
$^{4}$ \quad School of Astronomy and Space Science, University of Science and Technology of China, Hefei~230026,~China\\
$^{5}$ \quad University of Chinese Academy of Sciences, 19 A Yuquan Rd, Shijingshan District, Beijing~100049,~China}

\corres{Corresponding authors: yaoyh@cqu.edu.cn (Y.-H.Y.); guoyq@ihep.ac.cn (Y.-Q.G.)}

\abstract{The spatial-dependent-propagation (SDP) model with a nearby source works well to reproduce the coevolving features of both cosmic-ray (CR)-nuclei spectra and anisotropy. However, it is well known that the Sun is actually deviating from the galactic disk. This will lead to a dominating anisotropy in the direction perpendicular to the galactic disk, which is discrepant with current observations. Thus, it is necessary to further investigate the effect of the solar offset on anisotropy. In this work, {to the best of our knowledge, this is the first time that} 
 the combined studies of the solar offset, nuclei spectra, and anisotropy have been performed based on the SDP model. As a result, to reproduce CR spectra and anisotropy, the thickness of the inner halo $ (Z_{IH})$ needs to increase linearly with the displacement of the Sun. We also know that the PeV anisotropy could be used to estimate the value of the diffusion coefficient, thus breaking the degeneracy between the diffusion coefficient and halo thickness. Therefore, it is a good approach to constrain the halo thickness. Moreover, the anisotropy in the PeV energy region, as a new probe, might also shed new light on constraining the solar offset. It is hoped that the anisotropy of the energies from $\rm \sim$TeV to $\rm PeV$ can be finely measured with the LHAASO experiment, leading to a better understanding of the thick halo.}


\begin{document}

\section{Introduction}

In recent years, great progress has been made on the spectral measurement of cosmic rays (CRs) with balloon-borne and space-borne experiments. The fine structure of spectral hardening of nuclei at 200 GV was observed by ATIC-2 \citep{2006astro.ph.12377P}, CREAM \citep{2010ApJ...714L..89A}, and PAMELA \citep{2011Sci...332...69A}. AMS-02 also confirmed the hardening with unprecedented precision, although the spectral shapes have slight discrepancies from each other \citep{2015JCAP...09..023G}. At higher energies, the DAMPE observation clearly revealed that the proton spectrum further experiences a spectral softening at $\rm \sim$13.6 TeV, with the spectral index changing from $\rm \sim$2.60 to $\sim$2.85 \citep{2019SciA....5.3793A}. Hints of such spectral features were also found previously in CREAM \citep{2017ApJ...839....5Y} and NUCLEON measurements \citep{2018JETPL.108....5A}. These observations have stimulated significant concern among theoretical physical scientists. Some solutions that have been proposed to explain the spectral features include the contribution from nearby SNRs \citep{2012MNRAS.421.1209T}, the reacceleration mechanism of old SNR sources \citep{2010ApJ...725..184B,2014A&A...567A..33T}, the combination effects from different group sources \citep{2006A&A...458....1Z,2011PhRvD..84d3002Y}, and the spatial-dependent propagation (SDP) of CRs \citep{2012ApJ...752L..13T,2015ApJ...815L..25G,2016ChPhC..40a5101J}.

Although the structures of the spectra and anisotropy could be explained, one critical factor has been ignored in previous works, i.e., the Sun's offset. Usually, the solar system is assumed to be located at the midplane of the galactic disk, and the source distribution is symmetric above and below the disk. Yet, it has long been known that the Sun is located slightly above the galactic plane (toward the north galactic pole). The inferred distance above the midplane is from several to $\rm \sim$30 pc \citep{2007MNRAS.378..768J,2016AstL...42..182B,2017MNRAS.468.3289Y}. The offset would induce a net vertical flow outward from the galactic plane, which generates a corresponding anisotropy component. This is totally inconsistent with experimental observations. However, the estimated solar offset has a significant spread of values when employing a variety of different methods, although recent analyses have revealed smaller uncertainties. The combined study of nuclei spectra and anisotropy may shed light on this topic. In this work, further investigation aimed at the anisotropy problem is performed.

The rest of this paper is organized as follows. In Section~\ref{sec.2}, the model is described, and the calculated results compared with observations are presented in Section~\ref{sec.3}. Finally, conclusions are given in Section~\ref{sec.4}.


\section{Model Description}\label{sec.2}
Supernova remnants (SNRs) are recognized as significant accelerators of galactic CRs, capable of reaching energies up to a few PeV \citep{1978MNRAS.182..147B,1978MNRAS.182..443B,1978ApJ...221L..29B}. Before arriving at Earth, those CRs have traveled in the galaxy for $\rm \sim$$10^7$ years after diffusing away from the acceleration sites \citep{1977ApJ...217..859G}. During the journey, the impacts due to fragmentation and radioactive decay in the interstellar medium (ISM) result in the production of secondary particles. Additionally, the electrons suffer energy loss in the interstellar radiation field (ISRF) and magnetic field. This journey can be described by the propagation equation as
\begin{equation}
\begin{array}{lcll}
\frac{\partial \psi{(\vec{r},p,t)}}{\partial t} &=& {\rm q}(\vec{r}, p,t) + \vec{\nabla} \cdot
\left( D_{xx}\vec{\nabla} \psi - \vec{V_{c}} \psi \right)\\
&+& {\rm \frac{\partial}{\partial p}}[p^2D_{pp}\frac{\partial}{\partial p}\frac{1}{p^2}\psi]
- \frac{\partial}{\partial p}\left[ \dot{p}\psi - \frac{p}{3}
\left( \vec{\nabla}\cdot \vec{V_c}\psi \right) \right]\\
&-& \frac{\psi}{\tau_f} - \frac{\psi}{\tau_r},
\end{array}
\label{CRsPropagation}
\end{equation}
where ${\rm q}(\vec{r}, p,t)$ is the acceleration source, $\psi(\vec{r},p,t)$ is the density of CR particles per unit momentum $p$ at position $\vec{r}$, $\vec{V_c}$ is the convection velocity, 
$ \dot p\equiv dp/dt$ is the momentum loss rate, and $ \tau_f$ and $\tau _r$ are the characteristic timescales for fragmentation and radioactive decay, respectively. $ D_{xx}$ and $D_{pp}$ are the diffusion coefficients in coordinate and momentum space, respectively. For simplicity, the process of convection are not considered in this work.


\subsection{Spatial-Dependent Diffusion}
Located almost in the galactic disk, the galactic astrophysical objects would generate {large irregularities in the form of turbulence }
and affect the properties of particle diffusion. In the SDP model, the diffusion coefficient is assumed to be anticorrelated with the CR source distribution \citep{2012ApJ...752L..13T,2018PhRvD..97f3008G}; the diffusive volume is divided into two regions as an inner halo (IH,~$ |z| \leq Z_{\rm IH}=\xi Z_{\rm H}$) and an outer halo (OH, $ |z| > Z_{\rm IH}$). Here, $Z_{\rm H}$ is the half-thickness of the diffusive halo, and $\xi$ is the ratio of the inner halo to the entire halo with a typical value of ~$\rm\sim$0.1. In the IH region, the level of turbulence is expected to be high due to supernova-explosion activities; hence, the diffusion coefficient is relatively small. The OH region contains scarce active sources; thus, particles diffuse much faster. The diffusion coefficient is parameterized as \citep{2018PhRvD..97f3008G, 2018ApJ...869..176L}
\begin{equation}
D_{xx}(r, z, \mathcal R) = D_0 F(r, z) \beta^{\eta} \left(\frac{\mathcal R}{\mathcal R_0} \right)^{\delta_{0} F(r, z)} ~,
\end{equation}
where
\begin{equation}
F(r,z) =
\begin{cases}
g(r,z) +\left[1-g(r,z) \right] \left(\dfrac{z}{Z_{\rm IH}} \right)^{n} , &  ~|z| \leq Z_{\rm IH} \\ \rm~
1 ~, &~ |z| > Z_{\rm IH},
\end{cases}
\end{equation}
\begin{equation}
g(r,z) = \frac{N_m}{1+f(r,z)},
\end{equation}
where $\beta$ is the particle's velocity in units of light speed, $D_0$ and $\delta_0$ are constants representing the diffusion coefficient and its high-energy rigidity dependence in the outer halo, and $\eta = 0.05$ is a phenomenological constant in order to fit the low-energy data. The $N_m$ is a constant factor reflecting the property of irregular
turbulence in the halo. The constant $n$ describes the smoothness of the parameters at the transition between the two halos. $ f(r,z)$ is the source-density distribution. The numerical package DRAGON \citep{2008JCAP...10..018E} was used to solve the transport equation. In this work, the diffusion--reacceleration model has been adopted. The momentum-diffusive coefficient, $D_{pp}$, correlates with $D_{xx}$ via $D_{pp}D_{xx} = \frac{4p^{2}v_{A}^{2}}{3\delta(4-\delta^{2})(4-\delta)}$, where $v_A$ is the Alfv\'en velocity, $p$ is the momentum, and $\delta$ is the rigidity-dependence slope of the diffusion coefficient. The injection spectrum of background sources is assumed to be a power law of rigidity with a high-energy exponential cutoff, i.e., ${\rm q}({\cal R}) \propto {\cal R}^{-\nu} e^{(-{\cal R}/{\cal R}_{\rm c})}$. The cutoff rigidity of each element could be either $Z$- or $A$-dependent. The spatial distribution of sources takes the form of the SNR distribution \citep{1996A&AS..120C.437C}, $ f(r,z) \propto (r/r_\odot)^{1.69} e^{[-3.33(r -r_\odot)/r_\odot]} e^{(-|z|/z_s)}$, where $ r_\odot = 8.5$ kpc and $ z_{s} = 0.2$ kpc.

\subsection{Local Source}

The time-dependent propagation of CRs from the local source is obtained using the Green's function method, assuming a spherical geometry with infinite boundary conditions. The solution is
\begin{equation}
\phi(r, {\cal R}, t) = \dfrac{q_{\rm inj}({\cal R})}{(\sqrt{2\pi} \sigma)^3} \exp \left(-\dfrac{r^2}{2\sigma^2} \right) ~,
\end{equation}
where $ q_{\rm inj}({\cal R})\delta(t)\delta(\textbf{r})$ is the instantaneous injection spectrum of a point source and \linebreak$ \sigma({\cal R}, t) =\sqrt{2D({\cal R})t}$ is the effective diffusion length within time $t$. The diffusion coefficient $\rm D({\cal R})$ is taken as the value near to the solar system. The injection spectrum is again parameterized as a cutoff power-law form, $ q_{\rm inj}({\cal R})=q_0{\cal R}^{-\alpha} e^{(-{\cal R}/{\cal R}'_{\rm c})}$. The normalization $\rm q_0$ and spectral index $\alpha$ are determined through fitting to the galactic-cosmic-ray (GCR) energy spectra and anisotropy data simultaneously.

\subsection{Solar Offset}\label{sec:solar1}
\begin{figure}[H]
\includegraphics[width=0.9\textwidth]{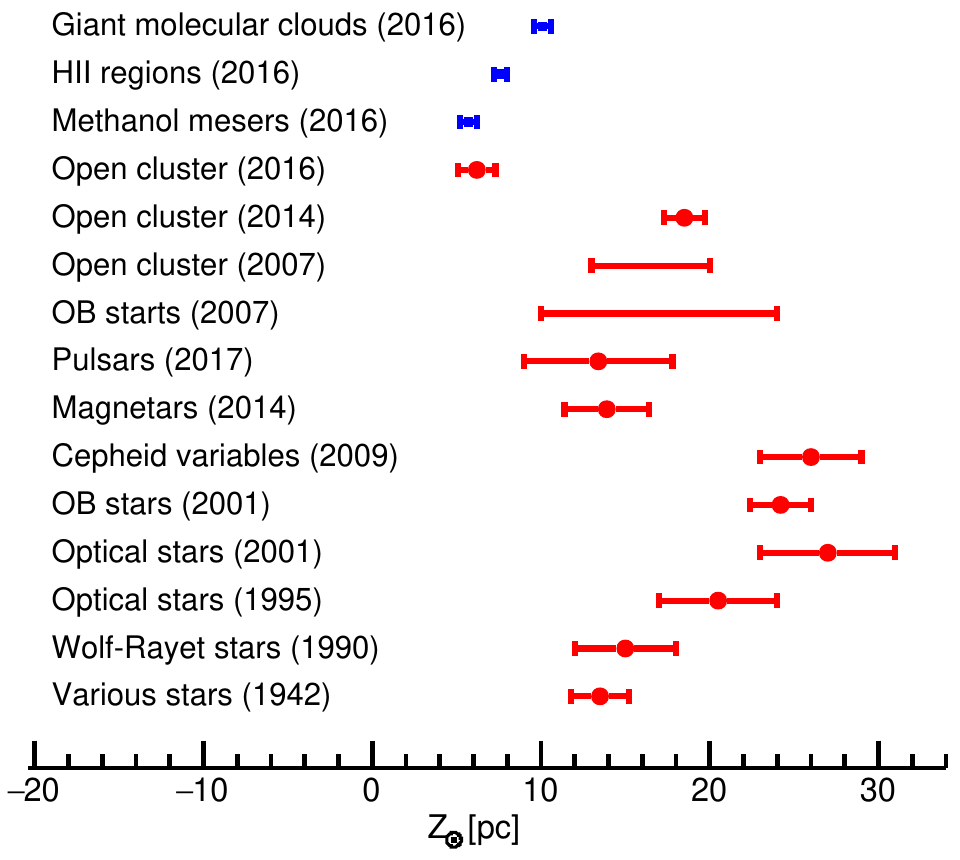}
\caption{Displacement of 
solar offset from the galactic plane summarized in \citet{2017MNRAS.468.3289Y}. Red points represent results from different star-based methods\citep{1942BAN.....9..315V,1990AJ....100..431C,1995AJ....110.2183H,2001ApJ...553..184C,2001AJ....121.2737M,2007MNRAS.378..768J,2009MNRAS.398..263M,2014ApJS..212....6O,2014MNRAS.444..290B,2016A&A...593A.116J}, and the three blue rectangles are results from \citet{2016AstL...42..182B}, obtained with nonstellar-astronomical-object methods.
}
\label{fig:solar}
\end{figure}

In the literature, several methods have been adopted to measure the solar offset from the galactic plane, which can be roughly classified into two categories: the matter-based ($\rm H_{II}$, molecular and methanol masers) and star-based (various {stars}
, pulsars, optical stars, Wolf--Rayet stars, cepheid variables, magnetars, and open clusters) methods. Figure~\ref{fig:solar} presents estimated offset distances, which are summarized in Table 1 of ref. \citep{2017MNRAS.468.3289Y}, from which it can be seen that the solar offset spans a large range, i.e., from $\rm5~pc$ to $\rm\sim$30~pc, although some analyses have small uncertainties. In the following, the effect of solar offset to the anisotropy is considered in the calculation.

\section{Results}\label{sec.3}
There are several components that might affect CR anisotropy that might remedy the nonconformity of anisotropy because of the above-inferred solar offset, namely, the source distribution, inner-halo thickness, and outer-halo thickness. At the beginning of this work, the effect of each component on the anisotropy was studied by fixing other components and taking the solar offset into account. We found that, except for the inner-halo thickness, the other two components have little influence on the anisotropy. In the following, the model results with a different $\rm Z_{IH}$ are mainly presented.

First, the influence of halo thickness on the CR flux density with the different solar offset is discussed. Then, the propagation parameters under various $\rm Z_{IH}$ {values }
are determined by fitting the ratio B/C, and the background component of proton spectra are obtained with the DRAGON package. Note that the contributions from the local source mentioned above are also included. Finally, the anisotropy amplitudes and their correlation with the vertical solar displacement from the galactic disk are given.

\subsection{Effect of Halo Thickness}
\par
\begin{figure}[H]
\includegraphics[width=0.65\textwidth]{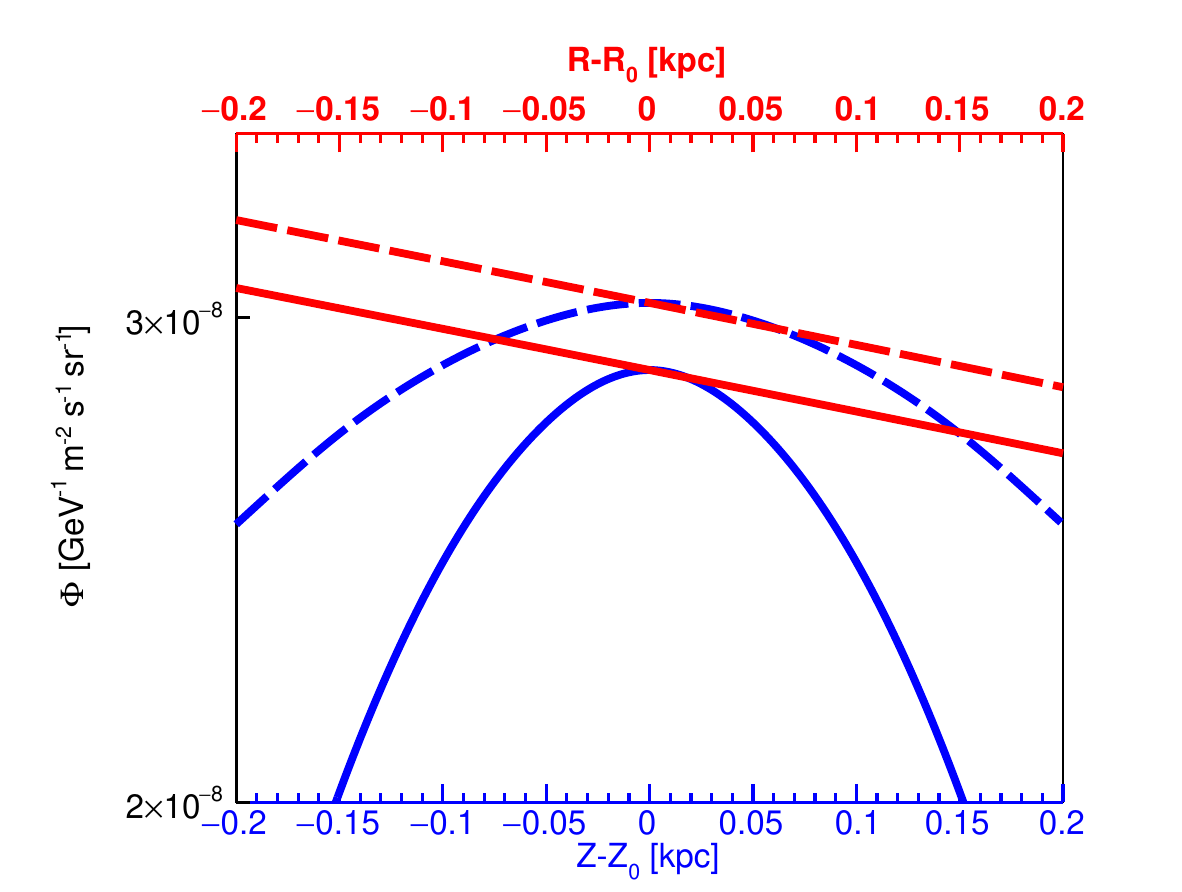}
\caption{
Schematic of 
CR fluxes along with distances from $\rm(R_0,Z_0)$ in vertical (blue) and radial (red) directions. Full (dashed) lines correspond with $\rm Z_{IH}=0.5~(1)$ and $\rm(R_0,Z_0)=(8.5, 0)~kpc$. Dashed lines have been scaled by 0.7 for convenience of comparison.
}
\label{fig:galaxy}
\end{figure}

As shown in Figure~\ref{fig:galaxy}, compared with the consistent radial anisotropy (slope of the CR flux), the anisotropy in the vertical direction changes significantly with an increasing vertical distance from the point $\rm(R_0,Z_0)$. Clearly, the thickness of the halo has a strong (little) influence on the vertical (radial) anisotropy. Therefore, a thicker halo would counteract the vertical CR stream caused by the solar offset, thereby reducing the total anisotropy.

\subsection{B/C and CR Spectra}
\par
\begin{figure}[H]
\includegraphics[width=0.65\textwidth]{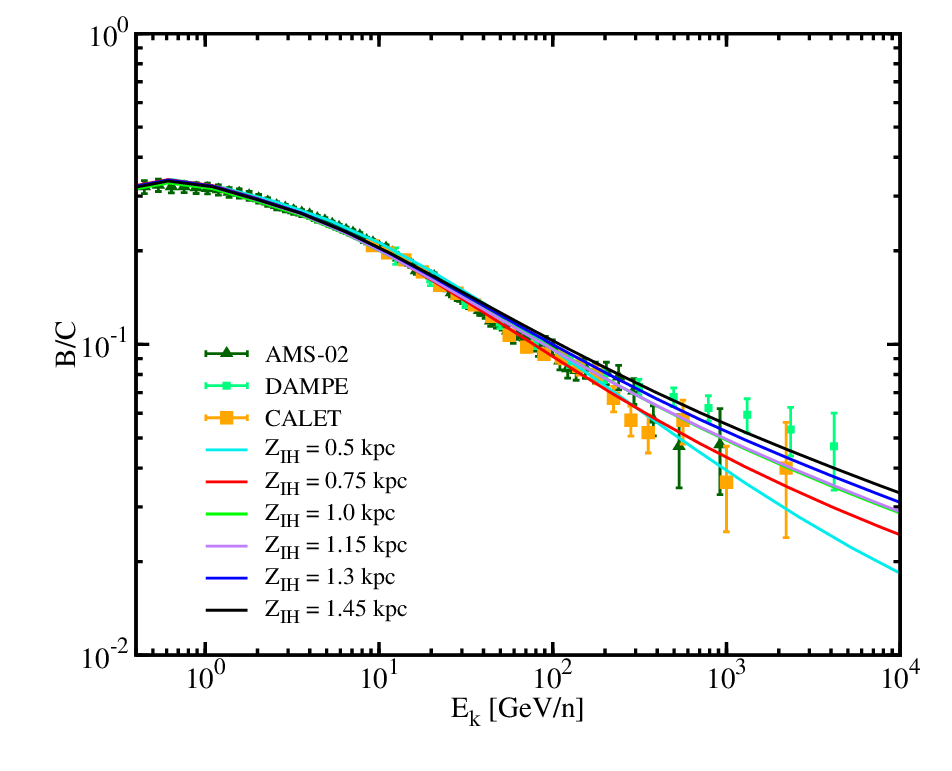}
\caption{\textls[-20]{Model predictions of B/C ratio compared with AMS-02, CALET, and DAMPE measurements~\citep{2015PhRvL.114q1103A,2017PhRvL.119y1101A,DAMPECOLLABORATION2022}}.}
\label{fig:bc}
\end{figure}

The transport parameters for the SDP model, with various values of $Z_{\rm IH}$, 
are determined based on the B/C ratio, as illustrated in Figure~\ref{fig:bc}. This figure compares the model intensities with data points from AMS-02 \cite{2015PhRvL.114q1103A}, CALET \cite{2017PhRvL.119y1101A}, and the latest data released from DAMPE \cite{DAMPECOLLABORATION2022}. The results show that the model fluxes, for different galactic-halo thicknesses, align well with the observations. Detailed parameters can be found in Table~\ref{tab:para}. The quantitative chi-square values are listed in Table \ref{tab:chi}.

\vspace{-12pt}

\begin{table}[H]
\caption{\label{tab:para}Propagation parameters $^\dagger$.}
\newcolumntype{C}{>{\centering\arraybackslash}X}
\begin{tabularx}{\textwidth}{CCCCC}
\toprule
$Z_{\rm IH}$ [kpc] &$Z_{\odot}$ [pc] &${D_0}$[$\times 10^{28}~{\rm cm}^{2}~{\rm s}^{-1}$] &  ${\delta}_0$ & $N_{\rm m}$\\
\midrule
0.5&   0  & .57  & 0.55 & 0.62\\
0.75& 5  &  6.9  & 0.65   & 0.39\\
1.0 & 10 &  8.75 & 0.65   & 0.39\\
1.15&  13 &  9.24 &0.69   &0.39 \\
1.3 &  16 & 10.07 & 0.69  &0.39 \\
1.45 & 19 & 11.05 &  0.7  &0.39\\
\bottomrule
\end{tabularx}
\noindent{\footnotesize{$^\dagger$ $\rm n$ and $\rm v_{A}$ are adopted as 3.5 and $\rm 6~km~s^{-1}$, respectively.
}}
\end{table}

\vspace{-12pt}
\begin{table}[H]	\caption{\label{tab:chi} Chi2/ndf for each distribution.}
\newcolumntype{C}{>{\centering\arraybackslash}X}
\begin{tabularx}{\textwidth}{CCCCCCC}
\toprule
$\mathbf{Z_{\rm IH}}$ \textbf{(kpc)}& \textbf{B/C} & \textbf{Proton}      & \textbf{Helium}       &  \textbf{All-Particle}    & \textbf{Aniso}  &  \textbf{Combined}\\
\midrule
0.5 & 255.84/99   & 203.17/112    &   238.95/105  &  13.63/33  & 3.42/3 & 715.01/352  \\
0.75&  91.96/99   & 406.94/112    & 134.08/105    & 19.76/33 &2.43/3 & 655.17/352 \\
1.0 & 106.25/99   &  295.72/112   & 134.78/105    &  24.2/33 &1.29/3 & 562.24/352  \\
1.15& 94.12/99    &  338.66/112   & 147.66/105     & 23.81/33 & 14.53/3 & 618.78/352\\
1.3 &  149.05/99  & 515.05/112    &  157.51/105      &  26.92/33 & 30.47/3 & 879/352  \\
1.45 &  233.72/99 &610.67/112    & 157.51/105     &  22.87/33  &50.79/3  &1069.94/352 \\
\midrule
data $^\dagger$ &    A(>15GeV)\linebreak+C1+D         &     A(>30GeV)+\linebreak C1+C2+D+K         &    A+C1\linebreak+C2+D+K              &     $\rm H\ddot{o}randel$      &    >PeV      &\\ 
\bottomrule
\end{tabularx}
\noindent{\footnotesize{$^\dagger$ \textls[-20]{The abbreviations A, C1, C2, D, and K correspond to AMS-02, CALET, CREAM, DAMPE and KASCADE-\linebreak Grande, respectively.}}}
\end{table}

\begin{figure}[H]
\includegraphics[width=0.5\textwidth]{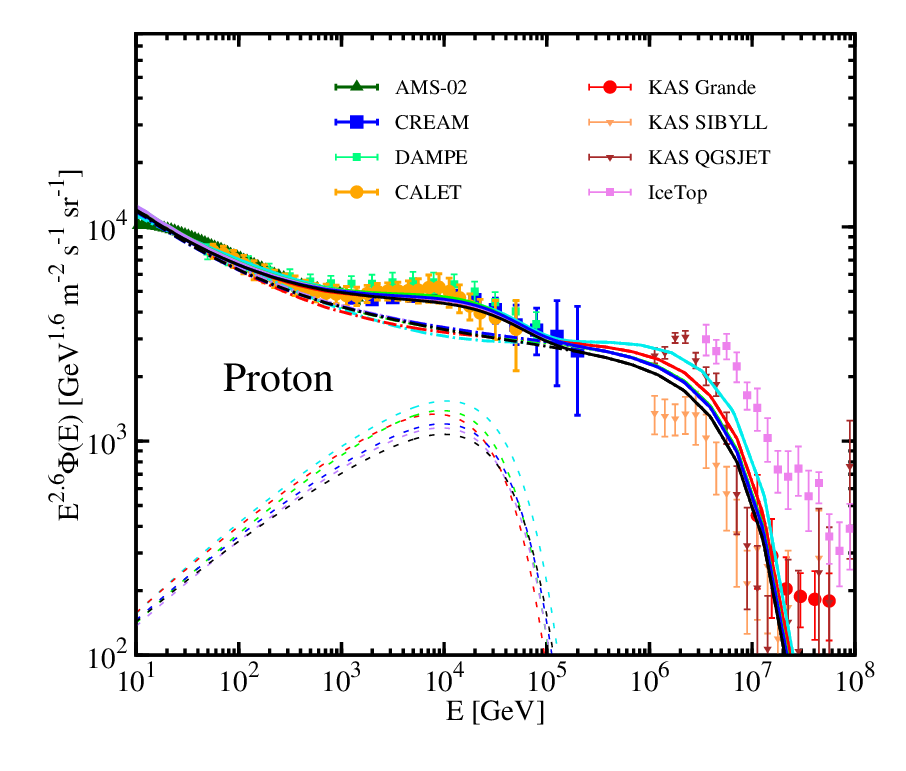}
\includegraphics[width=0.5\textwidth]{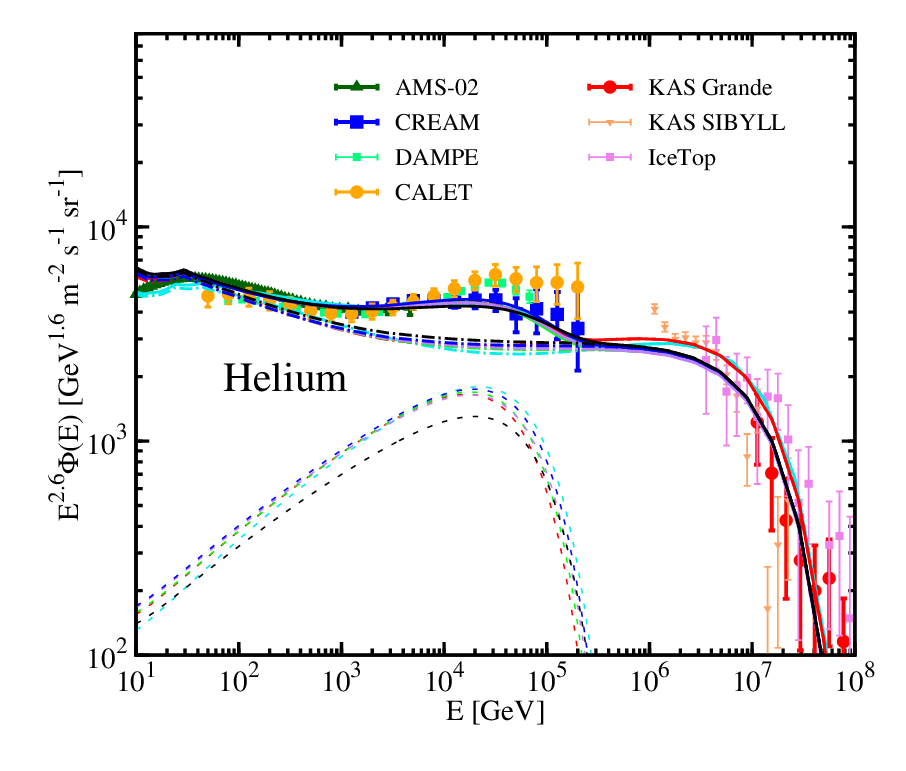}
\caption{Model predictions 
of protons and helium at different $\rm Z_{IH}$ values, compared with observations from AMS-02 \citep{2015PhRvL.114q1103A,2017PhRvL.119y1101A}, CALET \cite{2022PhRvL.129j1102A,2023PhRvL.130q1002A}, CREAM \citep{2010ApJ...714L..89A}, DAMPE \cite{2019SciA....5.3793A}, KASCADE \citep{2005APh....24....1A}, KASCADE-Grande~\cite{2013APh....47...54A}, and IceTop \cite{2019PhRvD.100h2002A}. Dotted-dashed and dashed lines are fluxes from background and local sources, respectively, and solid lines are their sums. Legends of lines are the same as Figure~\ref{fig:bc}, which indicate various $\rm Z_{IH}$ scenarios. 
}
\label{fig:proton}
\end{figure}

\begin{figure}[H]
\includegraphics[width=0.65\textwidth]{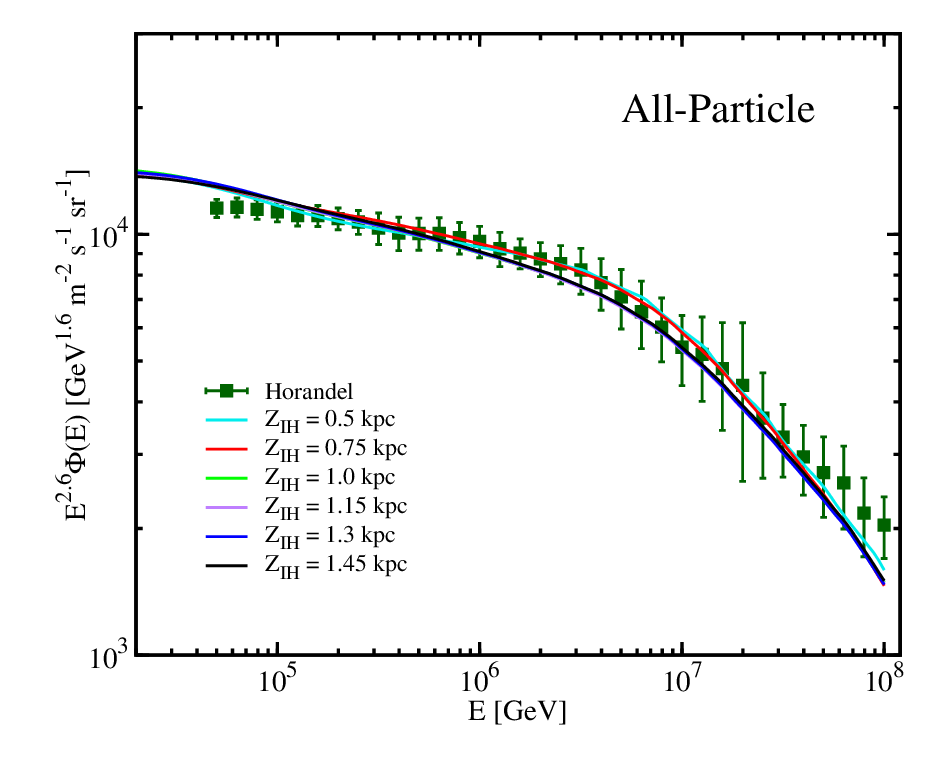}
\caption{
\textls[-20]{Model predictions of all-particle spectra at various $\rm Z_{IH}$ values, with the inclusion of the all-particle spectrum derived from extensive experimental data by \citet{2003APh....19..193H} for comparison.}
}
\label{fig:all}
\end{figure}

\vspace{-6pt}
After determining these fundamental parameters, we calculated and compared the propagated spectra of primary CR elements with experimental observations. For clarity and conciseness, we focused on presenting the proton and helium spectra, as depicted in Figure~\ref{fig:proton}. We utilized measurements from AMS-02 \citep{2015PhRvL.114q1103A,2017PhRvL.119y1101A}, CALET \cite{2022PhRvL.129j1102A}, CREAM \citep{2010ApJ...714L..89A,2023PhRvL.130q1002A}, DAMPE \cite{2019SciA....5.3793A}, KASCADE \citep{2005APh....24....1A}, KASCADE-Grande \cite{2013APh....47...54A}, and IceTop \cite{2019PhRvD.100h2002A} for comparison. Dotted-dashed and dashed lines are fluxes from background and local sources, respectively, and solid lines are their sums. The results indicate that the inclusion of a source component can simultaneously account for the spectral-hardening features at $\sim$200 GV and the softening features at $\sim$10 TV. It is worth noting that there are slight variations in the local-source component under different $\rm Z_{IH}$ scenarios to better fit the data. Furthermore, Figure~\ref{fig:all} presents the all-particle CR-energy spectrum predicted by the SDP model, alongside the $\rm H\ddot{o}randel$ all-particle spectrum derived from extensive experimental data \citet{2003APh....19..193H}. Table \ref{tab:chi} displays the chi-square values obtained from comparing model calculations with the observations. In the chi-square calculation process, the energy spectra of protons below 15~GeV and helium nuclei below 30~GeV are excluded due to solar modulation.

\vspace{-6pt}


\subsection{Anisotropy and Solar Offset}
\par

\begin{figure}[H]
\includegraphics[width=12cm]{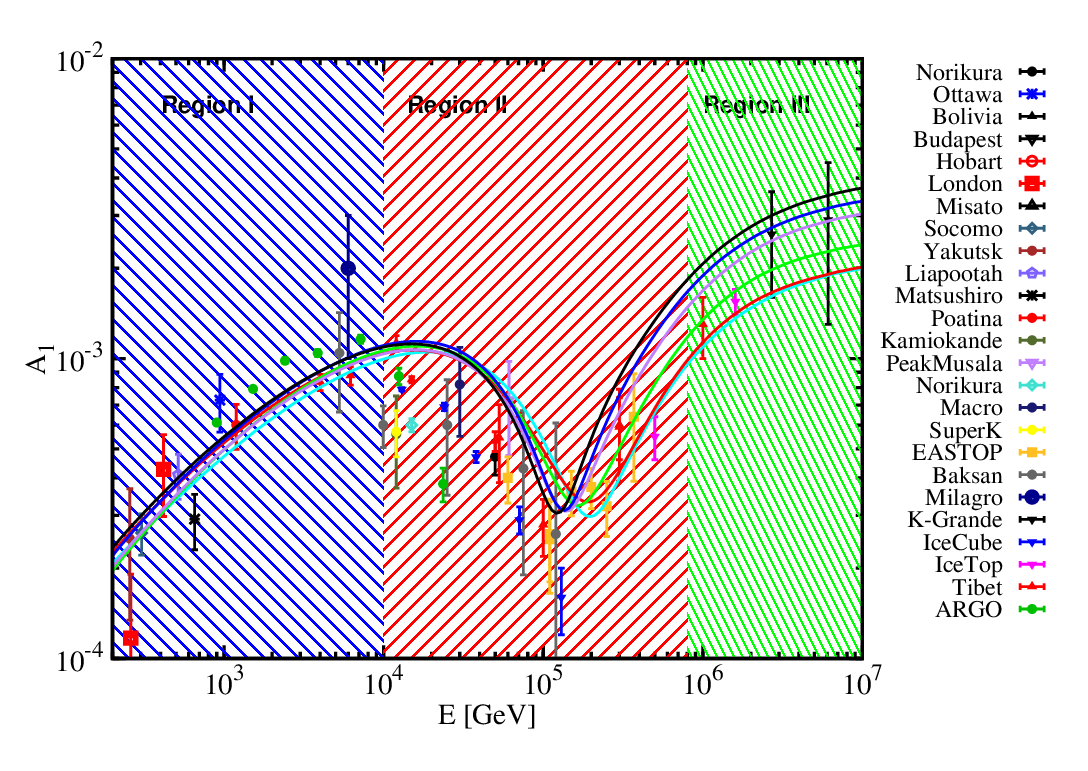}
\caption{Energy dependence of amplitudes of dipole anisotropy when adding all major CR elements together. Legends of lines are the same as Figure \ref{fig:bc}. The data points are taken from underground muon detectors: Norikura \cite{1973ICRC....2.1058S}, Ottawa\cite{1981ICRC...10..246B}, London \cite{1983ICRC....3..383T,1985P&SS...33.1069S}, Bolivia \cite{1985P&SS...33.1069S}, Budapest \cite{1985P&SS...33.1069S}, Hobart~\cite{1985P&SS...33.1069S}, Misato \cite{1985P&SS...33.1069S}, Yakutsk \cite{1985P&SS...33.1069S}, Banksan \cite{1987ICRC....2...22A}, Hong Kong \cite{1987ICRC....2...18L}, Sakashita \cite{1990ICRC....6..361U}, Utah \cite{1991ApJ...376..322C}, Liapootah~\cite{1995ICRC....4..639M}, Matsushiro \cite{1995ICRC....4..648M}, Poatina \cite{1995ICRC....4..635F}, Kamiokande \cite{1997PhRvD..56...23M}, Marco \cite{2003PhRvD..67d2002A}, and SuperKamiokande \cite{2007PhRvD..75f2003G}; and air-shower array experiments: PeakMusala \cite{1975ICRC....2..586G}, Baksan \cite{1981ICRC....2..146A}, Norikura \cite{1989NCimC..12..695N}, EAS-TOP \cite{1995ICRC....2..800A,1996ApJ...470..501A,2009ApJ...692L.130A}
Baksan \cite{2009NuPhS.196..179A}, Milagro \cite{2009ApJ...698.2121A}, IceCube \cite{2010ApJ...718L.194A,2012ApJ...746...33A}, Ice-Top \cite{2013ApJ...765...55A}, ARGO-YBJ \cite{2015ApJ...809...90B}, and Tibet \cite{2005ApJ...626L..29A,2015ICRC...34..355A,2017ApJ...836..153A}.}
\label{fig:aniso}
\end{figure}

It has been proposed in the literature that the energy-dependent anisotropy and softening features in the energy spectra may share a common origin \citep{2019JCAP...10..010L}. Figure~\ref{fig:aniso} demonstrates the successful reproduction of energy-dependent CR anisotropy amplitudes in this model, encompassing all major CR elements. By considering the contribution of the local source, the anisotropy of all components can be divided into three energy regions. \textls[-20]{In the lowest-energy region (region I), the anisotropy is primarily influenced by the propagation coefficient and local sources. At this stage, the former dominates, although the latter becomes increasingly important as the energy increases. The contribution of the local source to the anisotropy reaches its maximum value at tens of TeVs and then diminishes, playing a crucial role in reproducing the dip features observed in region II. In the highest-energy range (region III), the anisotropy increases as the galactic halo thickens. In this region, the effect of the local source becomes limited, and the propagation effect once again dominates the CR anisotropy.}

\begin{figure}[H]
\includegraphics[width=0.48\textwidth]{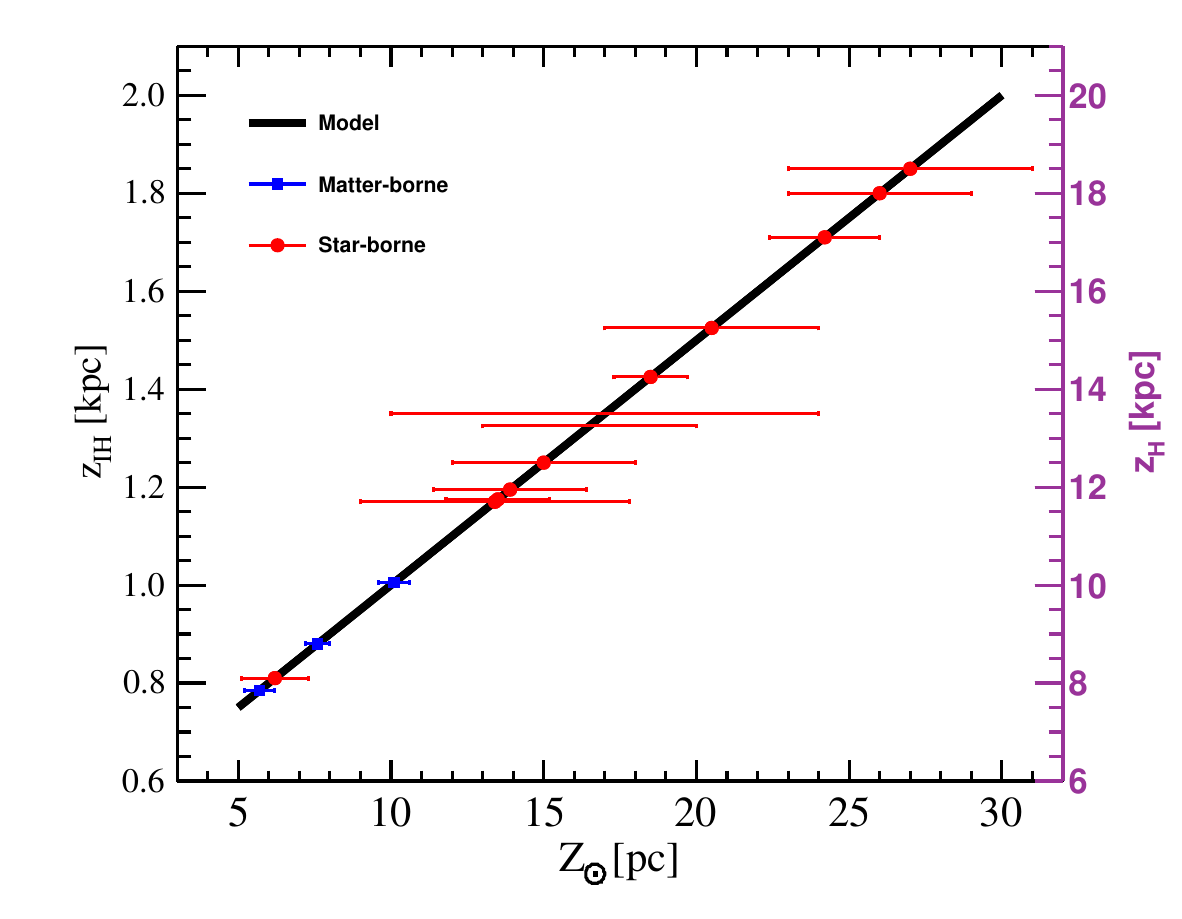}
\includegraphics[width=0.48\textwidth]{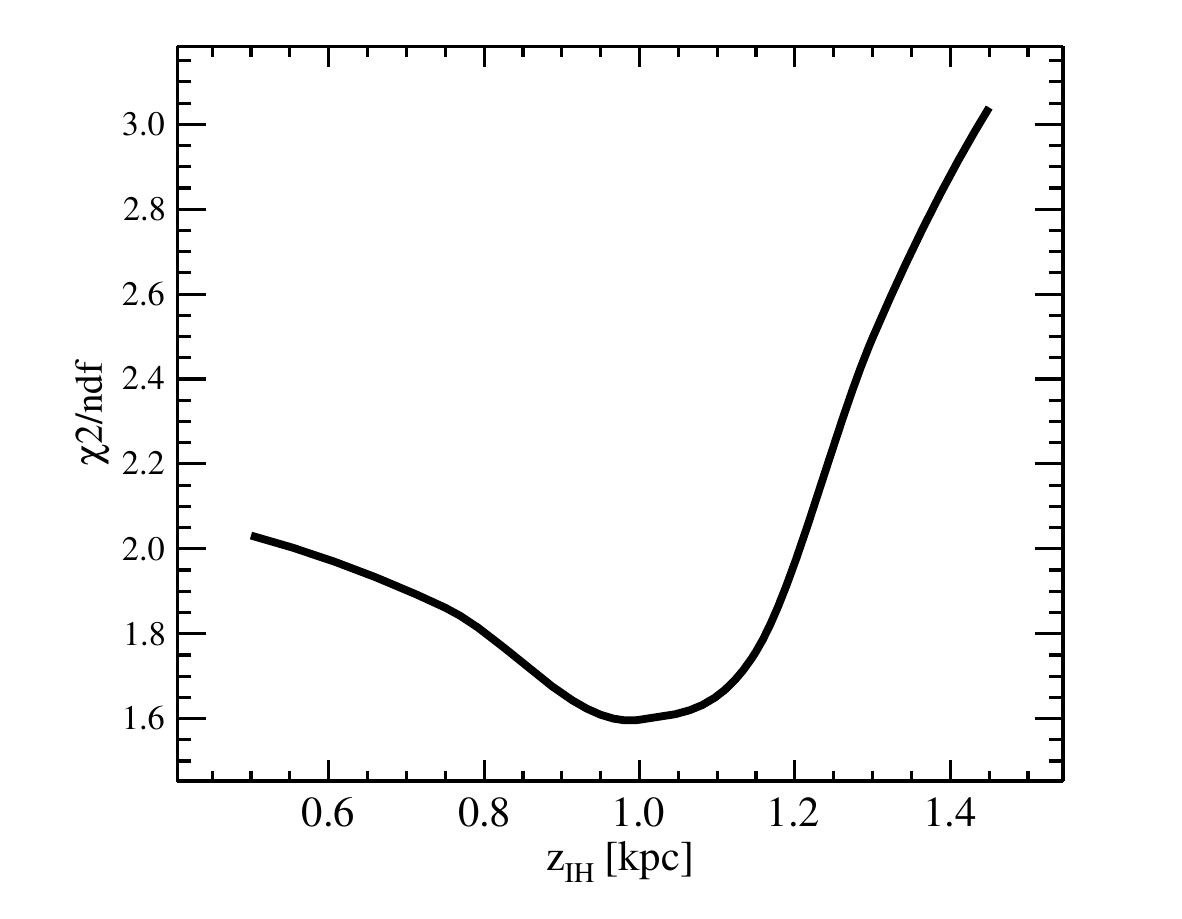}
\caption{Left: The black solid line represents the relationship between the displacement of the Sun from the point ($\rm R_0, Z_0$) in the galactic disk and the thickness of the halo. The right-hand axis also displays the total thickness of the halo, assuming $ \xi=\frac{Z_{\rm IH}}{Z_{\rm H}}=0.1$. The blue rectangles and red bullets represent the solar displacement estimated using matter- and star-borne methods, respectively. Right: The combined chi-square value varies with different values of halo thickness.}
\label{fig:inner}
\end{figure}

Owing to the limited contribution of the nearby source in the region III, the PeV anisotropy is a good approach to constrain both the halo thickness and the solar offset. As is known, the dipole anisotropy estimation is proportional to $\rm D_{xx}$ \citep{2012PhRvL.108u1102E}, while the ratio of the parameterized diffusion coefficient to the halo thickness, i.e., $\rm D_{xx}/Z_{\rm H}$($Z_{H}=Z_{\rm IH}+Z_{\rm OH}$), can be fixed by observing of the $\rm B/C$ ratio. Therefore, the thickness of the halo also affects the amplitudes of the anisotropy. That is to say, CR anisotropy is not only related to the diffusion coefficient $\rm D_{xx}$ but also to the thickness of the halo and the distance of the {Sun }
 deviating from the galactic disk. As previously mentioned, we have found through detailed analysis that only the inner halo has a significant impact on anisotropy. To~reproduce reversal structures of the dipole anisotropy, the thickness of the halo is required to increase linearly with the solar offset, as shown as in the left of Figure~\ref{fig:inner}. This figure also shows the values of halo thickness with the solar displacement measured with matter- and star-borne methods in Figure \ref{fig:solar}, under the relationship of halo thickness and solar offset estimated with PeV CR anisotropy.

To quantitatively assess the model's performance, we compared the calculated amplitudes of dipole anisotropy with observed values in region III. We utilized observational data points above PeV. The chi-squared values for different values of $\rm Z_{IH}$ (0.75, 1.0, 1.15, 1.3, and 1.45 kpc) were 2.4, 1.3, 14.5, 30.5, and 50.8, respectively, with a number of degrees of freedom equal to three. For the sake of simplicity, we temporarily disregarded the correlations between different energy spectra. Consequently, we aggregated the chi-square values of the~B/C~ratio, proton spectrum, helium spectrum, and all-particle spectrum, and the 4~PeV anisotropy points for the same set of model parameters, employing the additivity property of the chi-square distribution. The resulting total chi-square value is provided in the last column of Table \ref{tab:chi}. On the right side of Figure~\ref{fig:inner}, the combined chi-square value is depicted as it varies with different values of halo thickness. It is evident that the reduced chi-square value reaches its minimum when the halo thickness, $Z_{\rm IH}$, is approximately 1~kpc, which corresponds to a solar offset of 11 pc. When the halo thickness exceeds 1.2~kpc, the chi-square value increases sharply. Similarly, when the halo thickness is less than 0.8~kpc, the corresponding chi-square value also increases.

\section{Conclusions}\label{sec.4}

Measurements of CRs have entered a precise era, thanks to the fast development of space-borne and ground-based experiments in recent years. Based on the new features of CR spectra, including the spectral hardening at $\rm\sim$200~GV and softening at $\rm\sim$10 TV, together with the long-time enigma of the energy-dependent evolution of the dipole anisotropy features, an SDP framework with contributions from a local CR source was established and could explain most of these new observational facts \citep{2019JCAP...10..010L}.

{To the best of our knowledge, this work is the first time that the SDP model has been extended to simultaneously study the CR spectra, anisotropy of CRs, and the solar offset from the galactic disk}
. It was found that the primary CR protons, all particle spectra, and the dip structures of the CR anisotropy amplitude could still be well-reproduced after considering the actual solar location. At this moment, the thickness of the halo needs to increase linearly with the displacement of the Sun from the galactic disk because a thicker halo is required to counteract the vertical anisotropy due to the solar offset. Recent measurements, i.e., Fermi bubbles \citep{2010ApJ...724.1044S} and large-scale X-ray bubbles \citep{2020Natur.588..227P}, also support the point of a thick halo. Moreover, owing to the limited contribution of nearby sources in the PeV energy region, the anisotropy can be used to estimate the value of $ D_{xx}$, thus breaking the degeneracy of $\rm D_{xx}/Z_{H}$, which is tuned using the B/C ratio. Therefore, it is also a good approach to constrain the halo thickness. However, with present observations of CR anisotropy, our model shows that the solar offset prefers to be around 11 pc. It is worth noting that this might be a new (model-dependent) method of constraining the solar location. It is expected that future precise measurements of the anisotropy in the energy range from $\rm \sim$TeV to PeV, e.g., using LHAASO \citep{2019arXiv190502773B}, could provide a more accurate determination of the thickness of the halo, and, in addition, be used to test the proposed model and constrain the Sun's vertical location.

\vspace{6pt}



\authorcontributions{Conceptualization, B.-Q.Q., Y.-H.Y., and Y.-Q.G.; methodology, B.-Q.Q.; software, B.-Q.Q.; validation, W.L., Q.Y., X.-J.B., H.-B.H., and Y.-Q.G.; formal analysis, B.-Q.Q., Y.-H.Y., and Y.-Q.G.; investigation, B.-Q.Q.; resources, B.-Q.Q. and W.L.; data curation, B.-Q.Q.; writing---original draft preparation, Y.-H.Y.; writing---review and editing, Y.-H.Y. and Y.-Q.G.; visualization, Q.Y., X.-J.B., H.-B.H., and Y.-Q.G.; supervision, Q.Y. and Y.-Q.G.; project administration, Y.-Q.G.; funding acquisition, H.-B.H. and Y.-Q.G. All authors have read and agreed to the published version of the manuscript.}

\funding{This work is supported by the National Key $ R\& D$
Program of China (Grant No. \linebreak 2018YFA0404202) and the National Natural Science Foundation of China (Grant Nos. 11635011, 11875264, 11722328, 11851305, U1738205, and U2031110).}

\dataavailability {The experimental data used in this article are all from papers published by various experiments in official scientific journals. We have not created a dedicated dataset link yet. If anyone requires the relevant data, please contact the following email addresses: yaoyh@cqu.edu.cn; guoyq@ihep.ac.cn.
} 

\acknowledgments{This work 
is supported by the National Key $R\&D$ 
Program of China (Grant No. 2018YFA0404202) and the National Natural Science Foundation of China (Grant Nos. 11635011, 11875264, 11722328, 11851305, U1738205, and U2031110).}

\conflictsofinterest{The authors declare no conflicts of interest. The funders had no role in the design of the study; in the collection, analyses, or interpretation of data; in the writing of the manuscript; or in the decision to publish the~results.}

\sampleavailability{Samples of this article 
 are available from the authors.}


\abbreviations{Abbreviations}{
The following abbreviations are used in this manuscript:\\

\noindent
\begin{tabular}{@{}ll}
SDP &  spatial-dependent propagation\\
CRs & cosmic rays\\
GCR & galactic cosmic ray\\
SNRs & supernova remnants\\
ISM & interstellar medium\\
ISRF & interstellar radiation\\
outer halo & OH\\
inter halo & IH\\

\end{tabular}
}

\begin{adjustwidth}{-\extralength}{0cm}

\reftitle{References}

%


\PublishersNote{}
\end{adjustwidth}
\end{document}